\documentclass[twocolumn,prl]{revtex4}
\usepackage{amssymb}
\usepackage{graphicx}

\input{tcilatex}

\begin{document}

\title{Highly unperturbed inner tubes in CVD grown double-wall carbon
nanotubes}
\author{$^{1}$F. Simon}
\author{$^{2}$\'{A}. Kukovecz}
\author{$^{2}$Z. K\'{o}nya}
\author{$^{1}$R. Pfeiffer}
\author{$^{1}$H. Kuzmany}
\affiliation{$^{1}$ Institut f\"{u}r Materialphysik, Universit\"{a}t Wien, Strudlhofgasse
4, A-1090 Wien, Austria}
\affiliation{$^{2}$ Department of Applied \& Environmental Chemistry, University of
Szeged, Rerrich B. ter 1, H-6720 Szeged, Hungary}

\begin{abstract}
The synthesis of double-wall carbon nanotubes (DWCNTs) with highly
unperturbed inner shells is reported using the catalytic vapor deposition
method. Temperature dependent and high resolution Raman measurements show an
enhanced phonon life-time of the inner tubes with respect to the outer ones
and similar diameter SWCNTs. This proves that the inner tubes are
unperturbed similar to the inner tubes in peapod-grown DWCNTs. The presence
of the outer tube is argued to protect the inner tube from interaction with
impurities and also to stabilize the growth of defect free inner tubes. The
current material underlines the application potential of DWCNTs.
\end{abstract}

\maketitle







\section{Introduction}

Double-wall carbon nanotubes (DWCNTs) are on the borderline between
single-wall (SWCNTs) and multi-wall carbon nanotubes (MWCNTs). As a result,
they share common features with both classes of materials. Generally, SWCNTs
are considered to be of fundamental interest but limited for applications.
MWCNTs are known to be more application-friendly. SWCNTs exhibit a number of
compelling physical properties such as superconductivity \cite{TangSCI}, the
Tomonaga-Luttinger liquid state \cite{KatauraNAT,PichlerTLL} and a predicted
Peierls state \cite{BohnenPeierls,ConnetablePeierls}. DWCNTs are
fundamentally interesting as their physical properties are determined by the
well defined inner and outer tube chiralities \cite{PfeifferPRL,PfeifferEPJB}
and are candidates for applications such as e.g. reinforcing composites or
scanning microscopy probeheads due to the improved elastic properties as
compared to the SWCNTs but still of smaller diameter than MWCNTs. In
addition, the small diameter inner tubes enable to study the behavior of
highly curved nanostructures such as curvature induced deviations from the
electronic structure of graphite \cite{KurtiNJP2003,Simoncondmat0406343}.

DWCNTs can be grown from fullerenes encapsulated inside SWCNTs, peapods \cite%
{SmithNAT} (PEA-DWCNTs) by a high temperature treatment \cite%
{LuzziCPL2000,BandowCPL} without any catalysts. Raman studies on the inner
tubes in PEA-DWCNTs indicated at least an order of magnitude longer phonon
life-times, i.e. narrower line-widths, as compared to the outer tubes \cite%
{PfeifferPRL}. This was associated with the highly unperturbed nature of the
inner tubes, that are grown in the nano clean-room interior of SWCNTs. The
defect content of the tube shells is crucial for applications such as the
possibility of ballistic transport and improved mechanical properties.

An alternative DWCNT synthesis is based on catalytic SWCNT growth methods
such as arc-discharge \cite{HutchisonCAR} and Catalytic Chemical Vapor
Deposition (CCVD) methods \cite{ChengCPL,Lyu,Flahaut,WeiCPL,CiJAP,CiJPC}. So
far, Raman studies on such samples have given a similar linewidth for the
inner and outer tubes indicating comparable number of defects for the two
shells \cite{ChengCPL,WeiCPL,CiJAP,CiJPC,BacsaPRB}. However, no low
temperature and high resolution Raman data have been reported that is needed
for the accurate observation of the inner tube line-widths. As a result,
inner tubes in PEA-DWCNTs are the only tubes known to be highly defect free
among all kinds of CNT materials.

Here, we report that CCVD grown DWCNTs (CVD-DWCNTs) can also have very
narrow Raman line-widths indicating their high perfectness as the inner
tubes in PEA-DWCNTs. Interestingly, the inner tube diameter distribution in
the current CVD-DWCNT is very similar to the PEA-DWCNT samples and it is
discussed whether perfectness of small diameter inner tubes is a general
phenomenon of small diameter inner tubes.

\section{Experimental}

\textit{Catalyst preparation. }The CVD catalyst was a modified version of
the Fe/Mo/MgO system developed by Liu et al. \cite{LiuCPL} for SWCNT
synthesis. (NH$_{4}$)Mo$_{7}$O$_{24}$ (Reanal) and Fe(NO$_{3}$)$_{3}$%
\textperiodcentered 9 H$_{2}$O (Aldrich) were dissolved in distilled water
and the solution was added to the aqueous suspension of MgO and sonicated
for 30 minutes at room temperature. The resulting slurry was dried in two
steps, under vacuum at 90 
${{}^\circ}$%
C for 1 h and then at atmospheric pressure at 120 
${{}^\circ}$%
C overnight. The catalyst possessed a molar Fe:Mo:MgO ratio of 1:0.1:110.

\textit{\ Nanotube synthesis. }Nanotubes were synthesized in a fixed bed
horizontal quartz tube reactor with a diameter of 60 mm and a length of 110
cm. 0.30 g catalyst was placed in a quartz boat and shoved to the middle of
the reactor. After purging the system with Ar at room temperature, the gas
stream was changed to the C$_{2}$H$_{2}$:Ar (10 cm$^{3}$/min:150 cm$^{3}$%
/min volumetric flow rate at ambient temperature and pressure) reaction
mixture. The reactor was pushed into the furnace and kept there at 850 
${{}^\circ}$%
C for 20 minutes. Then, the reactor was removed from the furnace and purged
with a pure Ar stream until cooling down to room temperature. The catalyst
was removed from the sample by dissolving in excess of concentrated HCl
solution at room temperature. The remaining carbonaceous material was
filtered and washed with distilled water and dried at 120 
${{}^\circ}$%
C. Amorphous carbon content was diminished by oxidizing in flowing air at
300 
${{}^\circ}$%
C for 1 hour. The HiPco samples used as reference were purchased from CNI
(Carbon Nanotechnologies Inc., Houston, USA).

\textit{Raman spectroscopy.} Multi frequency Raman spectroscopy was
performed on a Dilor xy triple axis spectrometer in the 1.83-2.54 eV
(676-488 nm) energy range and in a Bruker FT-Raman spectrometer for the 1.16
eV (1064 nm) excitation at room temperature and at 90 K. The triple axis
spectrometer can be operated in subtractive and additive mode allowing to
take normal and high resolution spectra. The high resolution mode has a
spectral resolution of 0.4 cm$^{-1}$ at 1.83 eV excitation as determined
from the spectrometer response to the elastically scattered light.

\begin{figure}[tbp]
\includegraphics[width=\hsize]{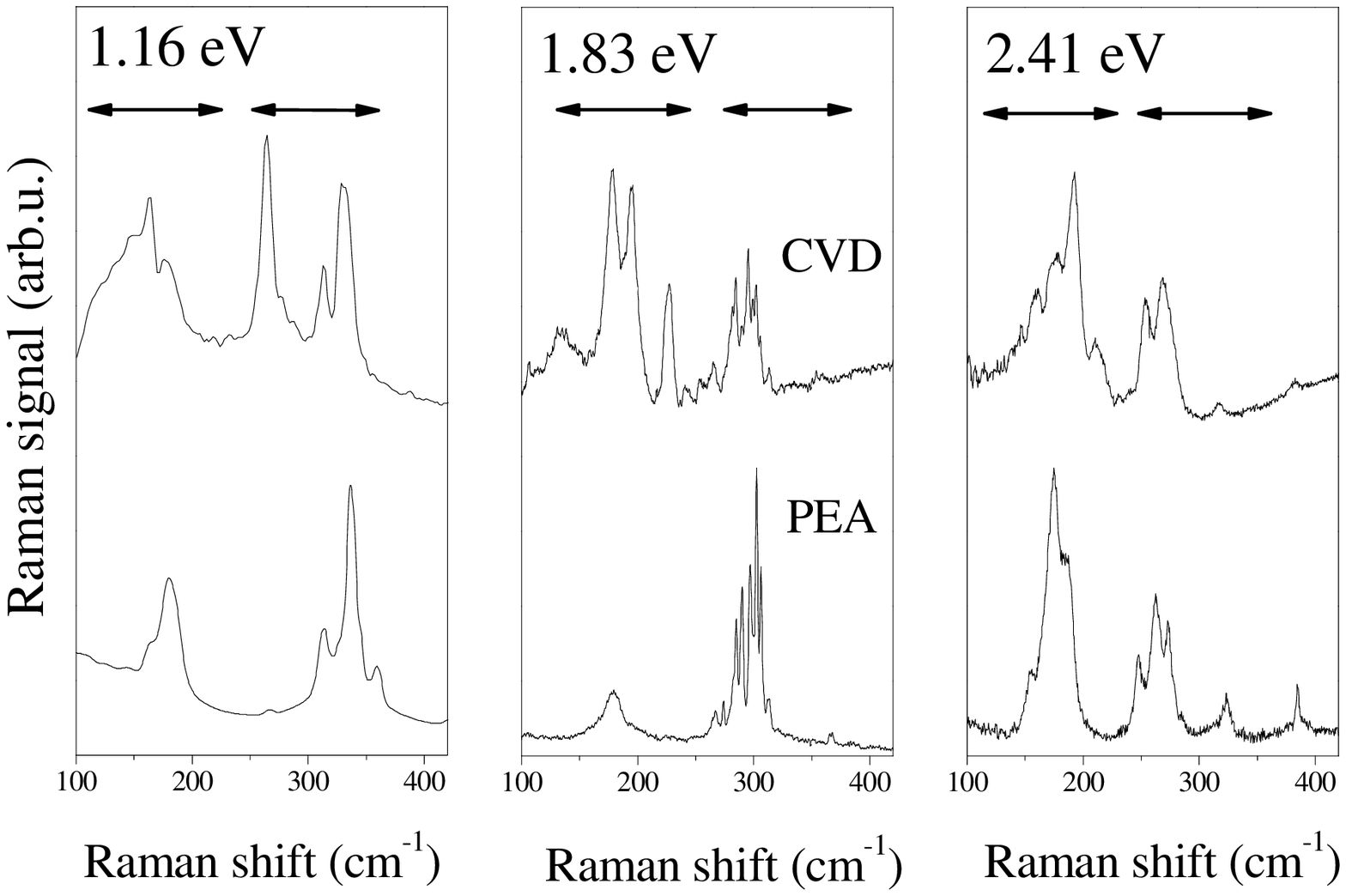}
\caption{Comparison of CVD- (upper curves) and PEA-DWCNT (lower curves)
samples at 1.16, 1.83, and 2.41 eV. The distinct lower and upper Raman shift
ranges indicated by arrows correspond to the outer and inner tube RBM
responses, respectively.}
\label{spectra}
\end{figure}

\section{Experimental results}

In Fig. \ref{spectra}., we show the Raman spectra of the CVD grown material
and the PEA-DWCNT samples for three laser excitations. Comparison of spectra
taken with different laser excitation energies is required due to the strong
photoselectivity of the nanotube Raman response. The radial breathing mode
(RBM) of SWCNTs with $d\approx $2.1 to 0.6 nm dominates the Raman response
for the displayed 120 to 400 cm$^{-1}$ spectral range. The spectral ranges
of 120-220 cm$^{-1}$ and 240-400 cm$^{-1}$ were previously identified with
the outer and inner tube Raman responses for the PEA-DWCNTs, respectively 
\cite{PfeifferPRL}. Surprisingly, the CVD grown material shows a similar
pattern with higher and lower Raman shifted ranges and a gap in between. We
also observed similar spectra for the two materials at other excitation
energies (not shown). It is tempting to assume that in the CVD grown
material the same spectral regions come from inner and outer tubes, too.
This means that the current CVD sample contains a sizeable amount of DWCNTs
rather than a broad distribution of SWCNTs as it was originally found for
tubes from a similar preparation \cite{LiuCPL}. Below, we present other
spectroscopic evidence which further supports that the lower and higher
spectral ranges originate from outer and inner tubes in the CVD-DWCNT
samples, respectively.

Although the inner and outer tube diameter distribution is very similar in
the two materials, the Raman intensities are different: the inner tube modes
Raman intensities are on average a factor two smaller for the CVD-DWCNT than
for the PEA-DWCNT sample. This indicates a smaller density of the
corresponding DWCNTs in the earlier sample, however, an accurate
determination based on the Raman intensities alone is not possible as it is
affected by a number of sample dependent factors such as the surface
morphology.

\begin{figure}[tbp]
\includegraphics[width=0.8\hsize]{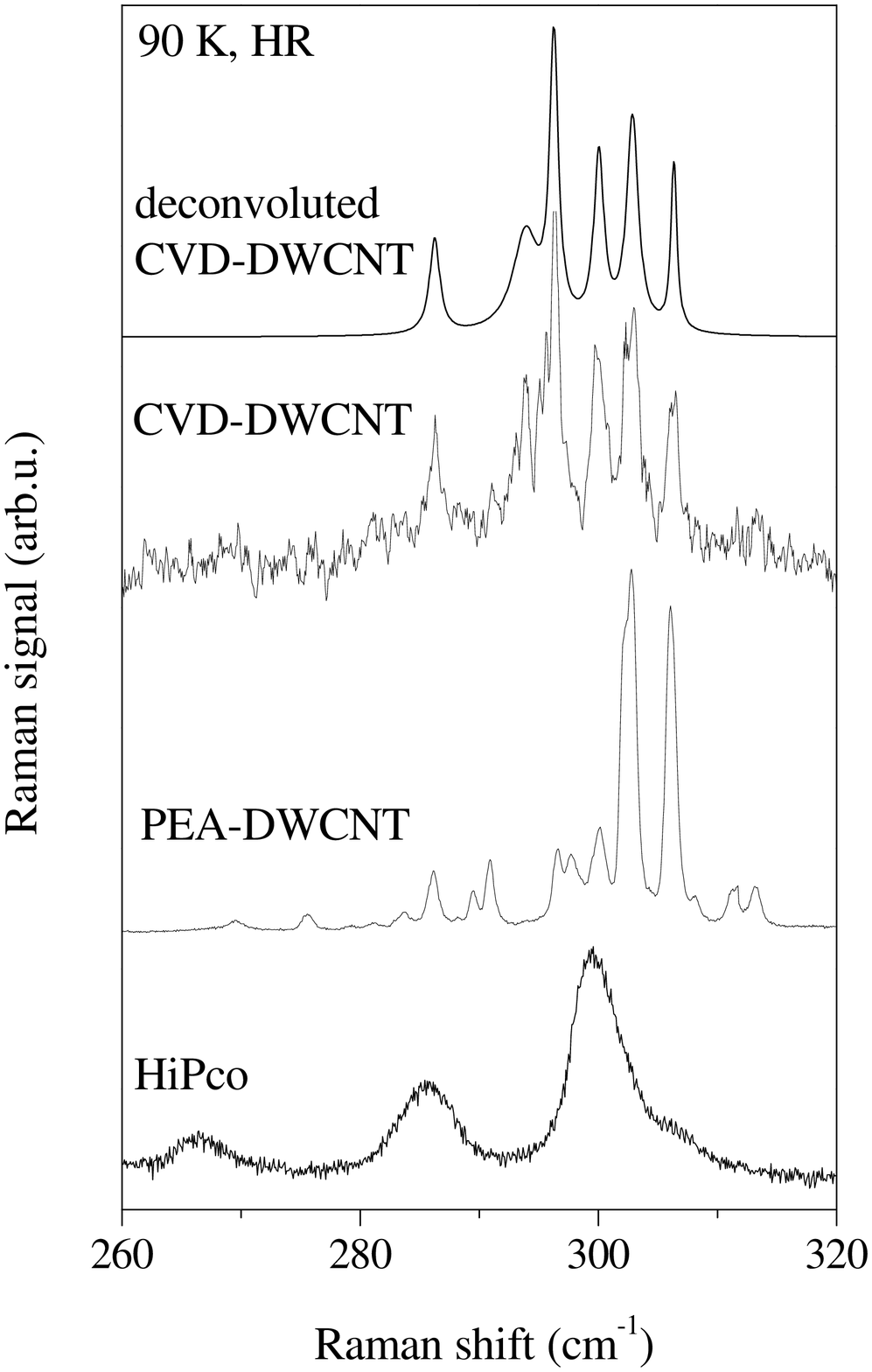}
\caption{High resolution Raman spectra taken at 1.83 eV excitation and 90 K
on the CVD- and PEA-DWCNT and an SWCNT reference (HiPco) sample. The
deconvoluted spetrum is also shown for the CVD-DWCNT sample. The narrow
linewidths indicate the long radial breathing mode phonon lifetimes of the
inner tubes in both DWCNT materials.}
\label{HRRamanspectra}
\end{figure}

The determination of the phonon life-times in the inner tubes in CVD-DWCNTs
is the important contribution of the current work. This parameter can be
used to characterize the defect content on the tubes. Often, the intensity
of the defect induced D-mode is used to estimate the defect concentration.
However, it was shown recently that the D-modes of inner and outer tubes
overlap, thus preventing a meaningful analysis unless $^{13}$C labelling of
the inner tubes is used to separate their response \cite{Simoncondmat0406343}%
. Long phonon life-times in the inner tubes can be measured in
high-resolution (HR) Raman experiments at 90 K and the corresponding spectra
is shown in Fig. \ref{HRRamanspectra}. For comparison, we show a spectrum
taken on an SWCNT sample prepared with the HiPco process (lowest curve in
Fig. \ref{HRRamanspectra}). The HiPco sample contains tubes with similar
diameters as the inner tubes in the DWCNT samples. Remarkably, the inner
tubes have a very narrow line-width for both DWCNT materials whereas the
line-width of the HiPco sample is significantly larger. The accurate
Lorentzian line-widths, that measure the intrinsic phonon life-times, were
determined by fitting the experimental spectra with Voigtian line-shapes
whose Gaussian component is the spectrometer response for the elastically
scattered light. The resulting deconvoluted spectrum is also shown for the
CVD-DWCNTs in Fig. \ref{HRRamanspectra}. Values for the FWHMs of the
Lorentzian lines down to 0.8 cm$^{-1}$ were found for the CVD-DWCNTs which
is almost as narrow as the FWHMs observed for the PEA-DWCNTs, down to 0.4 cm$%
^{-1}$ \cite{PfeifferPRL}. This is an order of magnitude smaller than the
line-widths of 5-6 cm$^{-1}$ found for the HiPco sample, for the outer tubes
in the current samples, or for individual SWCNTs \cite{JorioPRL2001}
indicating the highly unperturbed nature of the inner tubes in both
materials. The observable RBM\ modes agree in both materials within the 0.3
cm$^{-1}$ experimental precision of the line position measurement of the HR
experiment. The reason for the larger number of RBMs in the DWCNT samples as
compared to the HiPco sample was explained previously by the number of
possible inner and outer tube pairs. It was shown that an inner tube can be
grown in several outer ones and the corresponding RBMs are modified due to
the inner-outer tube interaction \cite{PfeifferPRL,PfeifferEPJB}. The
current result shows that the splitting is present irrespective of the DWCNT
growth method. This confirms that the origin of the splitting is the
inner-outer tube interaction as previously thought.

\begin{figure}[tbp]
\includegraphics[width=\hsize]{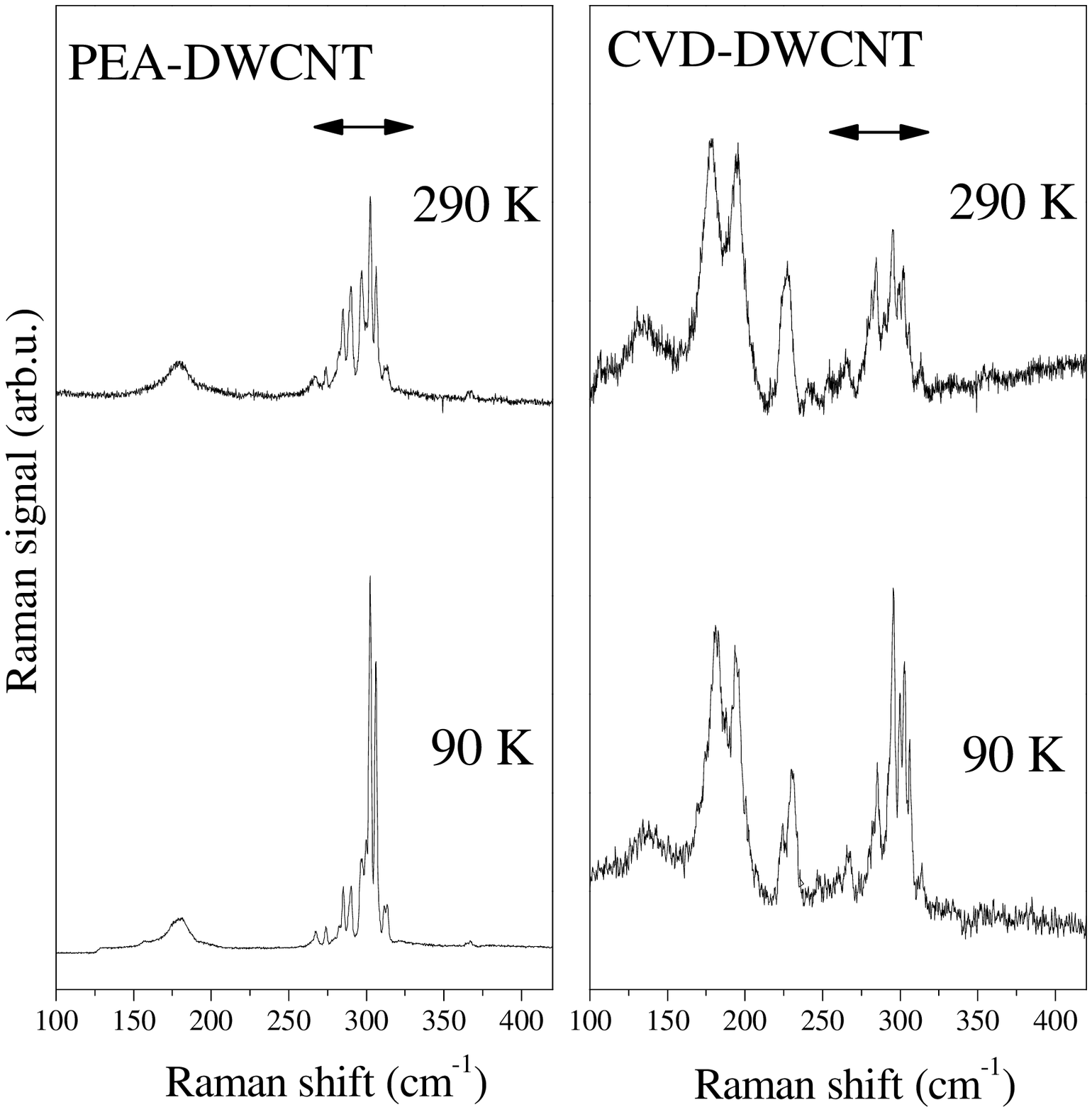}
\caption{Temperature dependence of the CVD- and PEA-DWCNT spectra taken at
1.83 eV excitation. The inner tube RBM modes (arrows) increase at lower
temperature compared to the outer tube RBMs.}
\label{Tdepspectra}
\end{figure}

In Fig. \ref{Tdepspectra}, we show the temperature dependent Raman spectra
for the two materials for the RBM ranges at the 1.83 eV excitation. The
integrated intensity of the inner and outer tube RBMs follows a distinct
temperature dependence for both compounds: it increases by a factor $\sim $2
at lower temperatures for the inner and is constant for the outer tubes.
Similar changes can be observed for the other laser excitations. The
increase of integrated intensity proves that it is not caused by the
lengthening of phonon life-times, i.e. narrowing of the modes, but it is due
to an increasing life-time of the quasiparticle state which is excited
during the Raman process. The details of this mechanism are studied and
published separately \cite{Simonunpub}. In brief, the quasiparticle
life-time is temperature independent and short for the outer tubes, whereas
it is longer and further lengthens with decreasing temperatures for the
inner tubes. In general, dirty and clean systems are characterized by
temperature independent short and longer temperature dependent life-times,
respectively. Analogous situations are encountered for the momentum
relaxation time in dirty and clean conductors \cite{AshcroftMermin}.
Therefore, the observed effect suggests that the outer and inner tubes
behave as dirty and clean systems, respectively. The cleanliness of inner
tubes is associated with the low concentration of lattice defects or
impurities and that it is well protected from the environment.

\section{Discussion}

The similarity between the observed RBM\ patterns in the CVD- and PEA-DWCNT
samples suggests that the 120-220 cm$^{-1}$ and 240-400 cm$^{-1}$ spectral
ranges originate from outer and inner tubes in the two samples,
respectively. The larger number of observed than geometrically allowed tubes
for the latter range is a further evidence supporting the DWCNT nature of
the CVD-DWCNT sample as it is characteristic for DWCNTs and originates from
the interaction of inner and outer tubes \cite{PfeifferPRL,PfeifferEPJB}. In
addition, RBMs in the two spectral ranges have very different properties for
both samples. Linewidths of the RBMs in the 240-400 cm$^{-1}$ spectral range
become narrower on lowering temperature, reaching an order of magnitude
smaller line-widths than the temperature independent line-widths of the RBMs
in the 120-220 cm$^{-1}$ spectral range. Similarly, Raman intensities of the
RBMs in the 240-400 cm$^{-1}$ spectral range increase on decreasing
temperature whereas the Raman intensity of the 120-220 cm$^{-1}$ spectral
range modes is temperature independent. This proves that the two spectral
ranges come from tubes with distinct properties and further supports the
above assignment. This also shows that inner tubes have long phonon and
quasiparticle life-times indicating that they are highly defect free and are
well protected from the environment.

The observation of long phonon- and quasiparticle life-times, and the
concomitant high perfectness of inner tubes in non peapod-grown DWCNT is
surprising. The high perfectness of inner tubes in PEA-DWCNT sample was
associated with the catalyst-free growth conditions and the perfectness of
the growth conditions as the inside of the host tubes can be considered as a
nano clean-room \cite{PfeifferPRL,Simoncondmat0404110}. As a result, only
inner tubes in PEA-DWCNTs were thought to be highly perfect. This paradigm
appears to be violated by the current CVD-DWCNTs. Previous results have
considered large diameter inner tubes with $d\approx 1.52$ nm where the
inner tubes had a large linewidth \cite{ChengCPL}. Interestingly, the inner
and outer tube diameters in the currently described CVD-DWCNTs are very
similar to the PEA-DWCNTs. Similar diameter inner tubes as those used here
were reported, however the temperature dependent effects were not studied 
\cite{WeiCPL,CiJAP,CiJPC,BacsaPRB}. Therefore, it cannot be ruled out that
perfectness of small diameter inner tubes is a general feature.

Unfortunately, the dominant contribution to the SWCNT RBM phonon and
quasiparticle life-times is not fully understood. It has been speculated
that SWCNTs strongly interact with attached impurities as well as contain a
number of defects, partly produced during synthesis or during purification.
Inner tubes are well protected from attached impurities. The effect of
attached impurities can also give rise to an inhomogeneous broadening. This
would explain why the similarly small diameter tubes in the HiPco samples
show significantly broader RBMs. Concerning lattice defects during growth,
it can only be speculated that the ensemble of the two shells of a growing
DWCNT energetically prefers the formation of a highly perfect inner tube.
Alternatively, it is also possible that the inner tubes are grown in the
CVD-DWCNT material only when the outer tube has been pre-formed. Then, the
outer tube provides a host template similarly as in the case of PEA-DWCNTs,
which enable the growth of highly defect-free inner tubes. Clearly, more
theoretical work is required to understand the growth processes of both
peapod based and CVD grown DWCNTs.

\section{Conclusion}

In conclusion, we reported the CCVD synthesis of DWCNTs with highly
unperturbed inner tubes similar to the inner tubes in peapod based DWCNTs.
This is a surprising result as for the latter material the inner tubes are
grown in a catalyst free nano clean-room environment. The diameter
distributions of the inner tubes are similar for both kinds of materials and
we suggested that the diameter of the inner tubes plays an important role
for their defect-free nature. The currently synthesized material is expected
to have improved properties compared to other DWCNTs and is a potential
candidate for applications such as conducting wires with increased ballistic
transport length or reinforcing element with improved mechanical properties.

\section{Acknowledgement}

We thank D\'{o}ra M\'{e}hn for her assistance in the nanotube synthesis.
This work was supported by FWF project P17345, by EU projects
BIN2-2001-00580, MEIF-CT-2003-501099 and by OTKA T046491 and F046361. \'{A}.
K. and Z. K. acknowledge support from a Zolt\'{a}n Magyary and a J\'{a}nos
Bolyai fellowship, respectively.

\end{document}